\documentclass[10pt]{article}
\usepackage[latin1]{inputenc}
\usepackage[T1]{fontenc}    
\usepackage[dvips]{graphicx} 
\usepackage{color,epsfig}
\usepackage[english]{babel}
\usepackage{vmargin}
\usepackage{amsmath}
\usepackage{amsfonts}
\usepackage{amssymb}
\usepackage{makeidx}
\usepackage{multicol}
\author{Eric Bavu, John Smith and Joe Wolfe}
\title{Torsional waves in a bowed string}
\date{}
\setmarginsrb{2cm}{2cm}{2cm}{2cm}{1cm}{1cm}{1cm}{1cm}
\usepackage{fancyhdr}
\begin{document}
\markright{entête}
\begin{center}

\begin{LARGE}
\textsc{}
\vspace{0.4cm}
\textsc{\textbf{Torsional waves in a bowed string}}
\end{LARGE}

\vspace{0.0cm}
\textbf{Eric Bavu, John Smith and Joe Wolfe} \footnote{Address for correspondence: J.Wolfe@unsw.edu.au       61-2-93854954}
\\
\begin{small}
Music Acoustics, School of Physics,
University of New South Wales,
Sydney 2052 Australia 

\end{small}

\end{center}

\vspace{0.4cm}
\begin{abstract}
Bowing a string with a non-zero radius exerts a torque, which excites torsional waves. In general, torsional standing waves have higher fundamental frequencies than do transverse standing waves, and there is generally no harmonic relationship between them. Although torsional waves have little direct acoustic effect, the motion of the bow-string contact depends on the sum of the transverse speed $v$ of the string plus the radius times the angular velocity $(r\omega)$. Consequently, in some bowing regimes, torsional waves could introduce non-periodicity or jitter to the transverse wave. The ear is sensitive to jitter so, while quite small amounts of jitter are important in the sounds of (real) bowed strings, modest amounts of jitter can be perceived as unpleasant or unmusical. It follows that, for a well bowed string, aperiodicities produced in the transverse motion by torsional waves (and other effects) must be small. Is this because the torsional waves are of small amplitude or because of strong coupling between the torsional and transverse waves? We measure the torsional and transverse motion for a string bowed by an experienced player over a range of tunings. The torsional wave spectrum shows a series of harmonics of the translational fundamental, with  strong formants near the natural frequencies for torsion. The peaks in $r\omega$, which occur near the start and end of the 'stick' phase in which the bow and string move together, are only several times smaller than $v$ during this phase. We present sound files of the transverse velocity and the rotational velocity due to the torsional wave. Because the torsional waves occur at exact harmonics of the translational fundamental and because of similarities in the temporal envelope, the sound of the torsional signal alone clearly suggests the sound of a bowed string with the pitch of the translational fundamental. However, the harmonics that fall near the torsional resonances are so strong that they may be heard as distinct notes.

\end{abstract}

\newpage 
\tableofcontents
\addcontentsline{toc}{section}{Introduction}

\vspace{3cm}
\listoffigures
\newpage

\section*{Introduction}

In one of the most important books ever published in music acoustics, Helmholtz \cite{1}  described what is now known as Helmholtz motion: the steady state motion of an idealised, one-dimensional, bowed string. Many of the studies of the bowed string since that of Helmholtz have studied departures from or complications to Helmholtz motion [2-11]. One consequence of the finite radius of the string is that the bow, which acts at the surface of the string and not at its centre, exerts a torque that produces torsional waves. The relationship between translational and torsional waves in the bowed string is the subject of the present study.\\

Waves approximating Helmholtz motion may be produced in a string by a bow to whose hairs rosin has been applied. Because of the rosin, the static friction between the string and bow may, during transients, be considerably greater than the kinetic friction. Figure 1 shows the motion schematically for a string bowed at a point $\frac{1}{\beta}$ from one end, where $\beta>1$. If the period is $T$, static friction maintains contact between the string and the bow for a time $T\times(1-\frac{1}{\beta})$ (the stick phase), and the string slides rapidly past the string in the opposite direction for a time $\frac{T}{\beta}$ (the slip phase). For a string of finite radius, the speed of the point of contact on the surface of the string is the combination $v + r\omega$ of the translational speed $v$ and the angular velocity $\omega$ due to the torsional wave, where $r$ is the radius of the string.\\

\begin{figure}[ht]
\begin{center}
\leavevmode
\parbox{8.7cm}{\centering \includegraphics[width=8cm]{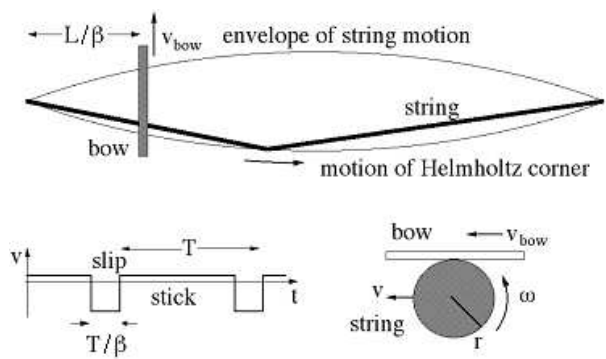}}\parbox{8.5cm}{\caption{\small{In ideal Helmholtz motion, the corner between two straight segments of a string (length $L$) traces two parabolæ in one period $T$. $v$ is the translational speed of the string, $\omega$ is the angular velocity due to the torsional wave  and $r$ is its radius. $v_{bow}$ is the speed of the bow.}}}
\label{fig1}
\end{center}
\end{figure} 
 
The fundamental frequencies of torsional waves in string instruments are rather higher than those of the translational waves. The speed of translational waves (and therefore the frequency) is proportional to the square root of the tension in the string: this dependence is used to tune strings. The speed of torsional waves depends on the square root of the ratio of torsional stiffness to moment of inertia, and depends only weakly on the tension \cite{12}. An important consequence of this different dependence on tensions is that there is, in general, no harmonic relation between waves of the two types. Gillan and Elliot \cite{13} studied torsional standing waves in strings using small coils attached to the string both for excitation and measurement of the waves. They showed that the ratio of the fundamental frequency of the torsional waves to that of the translational waves varied among strings of different pitch and of different manufacture.\\

The direct acoustic effect of torsional waves is small, because the torques that they exert on the bridge are small compared with those exerted by the translational waves. Nevertheless, as discussed above, the torsional waves could, in principle, have an important effect on the sound produced because they affect the motion of the contact point between bow and string. If the waves were in linear superposition, then one might expect that the varying relative phase between $v$ and $\omega$, and therefore that the variation from cycle to cycle in the value of $v+r\omega$, would lead to aperiodicity or jitter in the stick-slip motion. \\

A number of other effects give rise to small amounts of variation from one cycle to another, including those due to variations in the friction conditions and to the dynamics of the bow itself [7,8,11]. Human hearing is very sensitive to jitter \cite{14}, so even very small amounts are noticed, and they contribute to the sound of a real string instrument, as distinct from that made by a crude synthesiser. However, variations of a few percent would produce an unmusical sound. This is clearly not the case in a string bowed by an experienced player, who can reliably produce conditions that give rise to almost completely periodic, self-sustained oscillations. How is this achieved? Are the torsional waves small so that $r\omega\ll v$? Or are the torsional waves excited only at harmonics of the translational fundamental?\\

Torsional waves have been discussed by several authors [4,5,6,12],  Woodhouse and Loach \cite{12} measured their damping and their variation with string tension and Woodhouse et al.  \cite{11} discuss the effects of torsion in a study in which they determined the force at the bridge. The object of the present study is to measure the torsional and translational waves in a bowed string while it is being played, and to compare their amplitudes and phases. It aims to answer the questions: How large are the torsional waves produced during playing? and How are their frequencies related to those of the translational motion?


\section{Materials and Methods}

A string with high mass per unit length was chosen so that the mass of the sensors produced relatively little perturbation. The study was conducted using a string designed to play E1 (41 Hz) on  a bass guitar. It was steel cored, with three steel windings. Its mass per unit length was $20$ g.m$^{-1}$. Using a bass guitar machine head, it was stretched between two heavy, rigid 'bridges' , each consisting of a metal plate mounted on a hardwood block. The bridges were bolted to a brass frame mounted directly on a laboratory bench. The bridge separation was L $= 750$ mm. The tension was varied between experiments so that the transverse frequency took 13 different values over the range 30 to 60 Hz. The string has a flexural rigidity of $1.5$ mN.m$^{2}$, measured for static loads. (The flexural rigidity may increase with frequency \cite{5}.) Its torsional stiffness, measured in situ on the monochord at normal tension using a torsional pendulum, was $3.1$ mN.m$^{2}$.\\

A permanent magnet, of field strength 80 mT and approximately in the shape of a letter C, was used to produce the magnetic field, which was measured with a Hall probe mounted on a micro-manipulator. The pole pieces of the magnet were shaped by adding a series of rings whose thicknesses, inner and outer diameters and position were chosen empirically, by a process of iteration, to produce a region of homogeneous magnetic field. In the plane of symmetry between the poles, at right angles to the field, this process produced a region in which the field varied by less than $\pm1$\% over a circular region with radius 14 mm. The magnet was placed so that the bowed string passed through the centre of this region, and the direction of bowing was chosen to produce translational motion in this plane.\\

The position of the motion sensor is a compromise: if it is at a position $\frac{L}{n}$ from one bridge, where $n$ and $m$ are positive integers, then the $(n.m)^{th}$ harmonics are absent. So $n$ should ideally  be large. However, large $n$ means small amplitudes of both translational and torsional waves, and therefore an inferior signal:noise ratio. For thus study, $n = 5$.\\

To measure the translational motion, two segments of insulated copper wire, of length $12$ mm and diameter $100$ $\mu$m, were attached above and below the string as shown in Figure 2. At each end, these wires were bent into semicircular loops oriented so that the plane of the loop had no magnetic flux. The other ends of the loops were connected to preamplifiers. The average of the emfs produced by these two circuits is assumed to be proportional to the translational velocity of the string. The sum of the emfs in series was recorded on one channel of a digital oscilloscope at a sampling rate of 10 kHz and thence transferred as data to a computer. The relationship between this signal and the translational velocity was determined experimentally by driving the string with an electromagnetic shaker with an accelerometer head.\\

\begin{figure}[ht]
\begin{center}
\leavevmode
\parbox{8.7cm}{\centering \includegraphics[width=4cm]{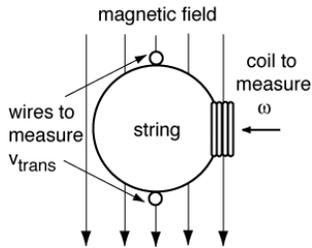}}\parbox{8.5cm}{\caption{\small{A sketch of the sensors mounted on the string. The sizes of the sensor wires have been magnified for clarity. The magnetic field is vertical so that the flux through the coil is zero when the string is at rest.}}}
\label{fig2}
\end{center}
\end{figure} 

In principle, the difference between the emfs of the two wires mentioned above would, in the small angle approximation, be proportional to the torsional emf. However, the signal:noise ratio of this signal was inadequate. Consequently, a coil of 10 turns, 12 mm long and 2 mm wide, made of insulated copper wire of 50 $\mu$m diameter, was attached to the string and oriented so that its magnetic flux was zero when the string was in mechanical equilibrium. The wires connecting the coil to another preamplifier were twisted, to reduce stray flux, and bent in a loop to minimise the force exerted on the string. The signal from this coil was recorded on the other channel of the oscilloscope and thence transferred to the computer. The flux linkage of this coil --- and hence the relationship between the signal and the angular velocity $\omega$ --- was measured by generating an oscillating field of known magnitude and frequency with a large coil. Fourier transforms were made using a Hann window.\\

The sensors were attached to the string with mylar tape. The total mass of the sensors was 20 mg, which is equivalent to the mass of a 1 mm length of the string. The force required to deform the connecting loops was orders of magnitude less than that required to displace the string laterally by the same distance, which was estimated by pushing them with a finger.\\

The fundamental frequency of the torsional motion, in the absence of a bow, was measured in a separate experiment using a method adapted from that of Gillan and Elliot \cite{12}. A driving coil (20 turns of 200 $\mu$m wire) was attached to the coil 15 mm from one end, and driven with an \textit{AC} current in the field of a permanent magnet. A sensing coil, 10 turns of 100 $\mu$m wire was attached 25 mm from the opposite end, in the field of another magnet. The frequency of the driving coil was varied to obtain a maximum in the signal of the sensing coil.\\

The Q of the fundamental resonances for translational and torsional standing waves was measured by giving the string respectively a translational and a rotational displacement at the bowing point and measuring the signals (low pass filtered to remove higher harmonics) as a function of time to obtain the decay time. The Q of translational standing wave fundamental of the E string on a double bass was measured (in situ) in a similar fashion.\\

The monochord string was bowed with a violin bow, chosen because it was less wide than a bass bow, and because the bowing position could therefore be specified more precisely. The bowing position, at a distance $\\frac{L}{beta}$ from one bridge, was varied between the sensor and the closer bridge over the range $\beta = 6$ to $15$, the first limit being the minimum possible distance between the bow and the sensor with this apparatus. The string was bowed in a horizontal direction, at right angles to the string. For the measurements shown here, it was bowed by the first author E.B., who had 17 years of experience playing violin. It was also bowed by three experienced bass players, according to whom, it was no more difficult to bow than a normal E string on a bass. For comparison, it was also bowed by people with little experience in bowing a string instrument (data not shown).


\section{Results and discussion}

For a translational wave on this string on a monochord, the characteristic decay time was typically 7.5 s. This corresponds to a Q value of typically 900. This value is rather higher than the value for a string on a double bass, because on the latter the bridge is less rigid and thus more energy is extracted from each cycle to drive the soundboard. For the E string on a double bass, the decay time was 0.45 s, corresponding to a Q of 57. For the torsional wave on the string on the monochord, the Q was 17.\\

\begin{figure}[ht]
\begin{center}
\leavevmode
\centering 
\includegraphics[width=14cm]{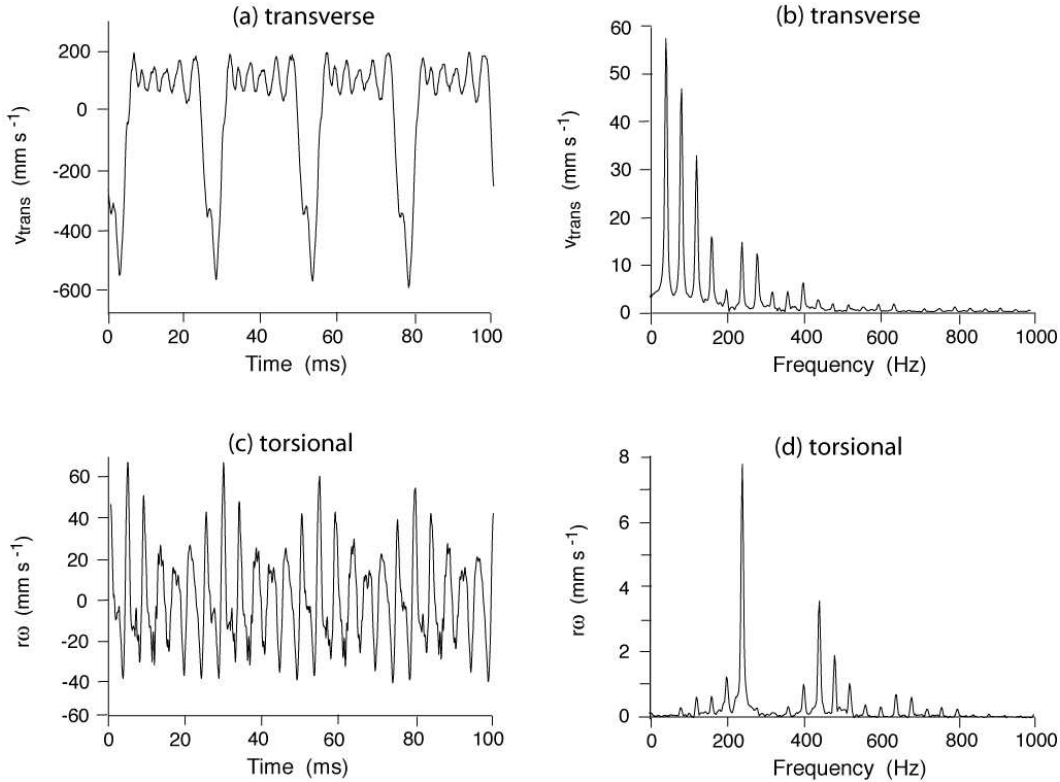}
\caption{\small{Time (left) and frequency domain (right) measurements of the translational (upper) and torsional velocity of the string, measured simultaneously on the string at point $\frac{L}{5}$. The string was bowed at $\frac{L}{6}$.}}
\label{fig3}
\end{center}
\end{figure} 

\begin{figure}[ht]
\begin{center}
\leavevmode
\centering 
\includegraphics[width=14cm]{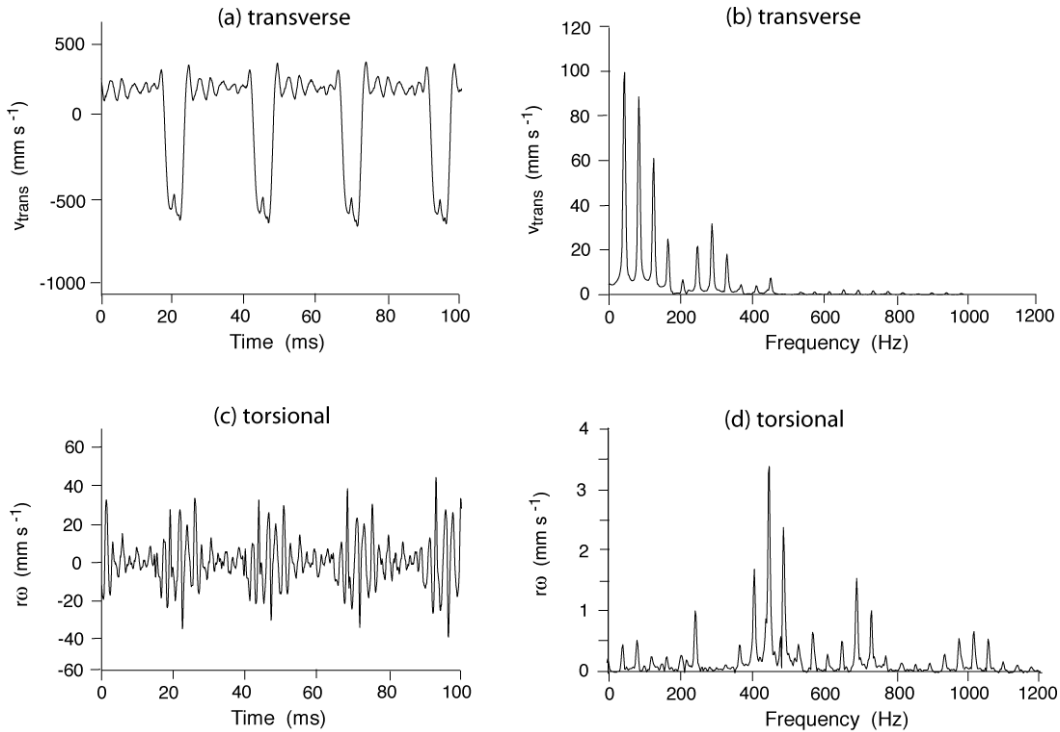}
\caption{\small{The string was bowed at $\frac{L}{13}$. The translational and torsional waves were measured simultaneously on the string at point $\frac{L}{5}$. The string was tuned so that the translational fundamental was 40 Hz. At this tension, the frequencies of the first two torsional resonances, measured in a separate experiment, were $240\pm2$ Hz and $472\pm3$ Hz.}}
\label{fig4}
\end{center}
\end{figure}

\begin{figure}[ht]
\begin{center}
\leavevmode
\centering 
\includegraphics[width=14cm]{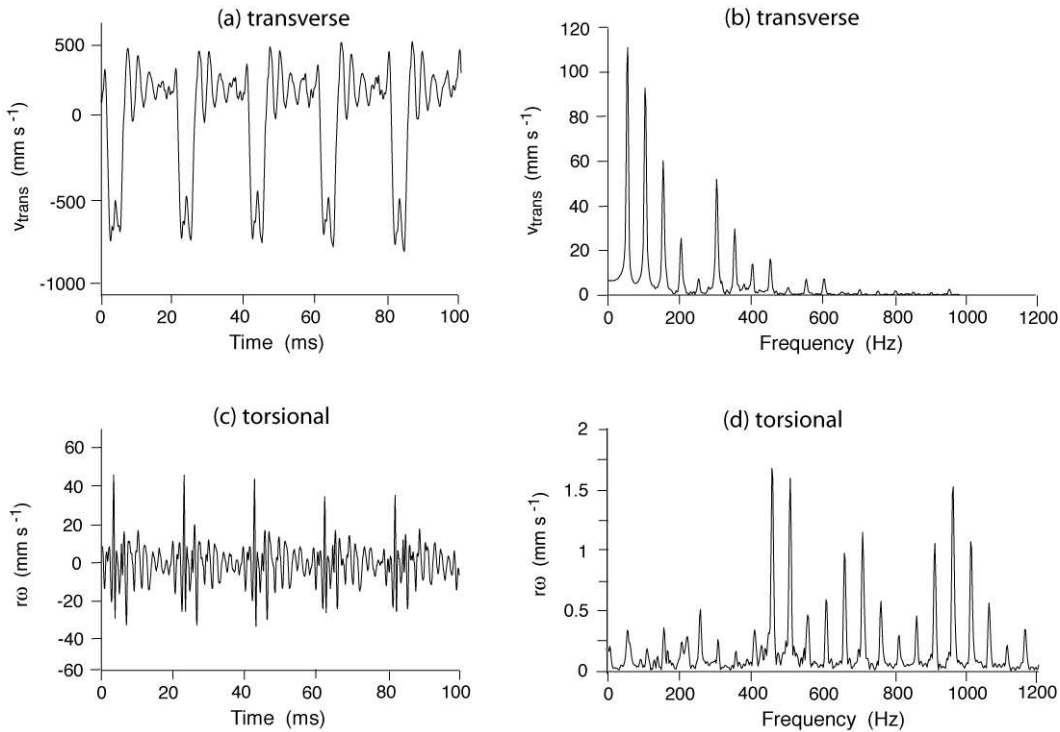}
\caption{\small{As in Figure 4, the string was bowed at $\frac{L}{13}$ and the translational and torsional waves were measured simultaneously on the string at point $\frac{L}{5}$. In this case, however, the string was tuned so that the translational fundamental was 50 Hz. At this tension, the first two torsional resonances were at $222\pm2$ Hz and $438\pm3$ Hz. }}
\label{fig5}
\end{center}
\end{figure}

Figures \ref{fig3}, \ref{fig4}, and \ref{fig5} show the translational and torsional waves produced when the string was bowed at different positions and at different tensions. In all cases, the results are given in the time domain at left, and in the frequency domain at right, for simultaneous measurements. The following features are common to results for all the bowing positions and tunings examined. The time domain representations of translational velocity show that the motion approximates Helmholtz motion: each period comprises a short phase, during which the string moves rapidly one way, and a longer phase during which the string moves more slowly in the opposite direction. The slow phase in all of the graphs shows several oscillations, which we discuss below. The motion is almost exactly periodic. The frequency domain also shows the approximation to Helmholtz motion. In ideal standing waves on a one dimensional string whose motion was measured (as here) at one fifth of its length, the fifth, tenth, fifteenth etc. harmonics would be zero. In all of the measurements, the fifth and tenth harmonics of the translational motion of the (centre of the) string were small but non-zero.\\

The harmonics above about the tenth are weaker than expected. The attenuation is unlikely to be produced by the finite bending stiffness and damping of the string because higher harmonics are present in the sound recorded by a microphone placed very near the string. We attribute this feature to high frequency attenuation in the measurement technique used to measure translational velocity. The finite length of the sensor is not alone great enough to explain the putative attenuation. However, it is possible that the finite mass of the sensor and the finite rigidity of the connecting loops might interact with the bowed string so as to attenuate the high frequency components.\\

The time domain representations of torsional velocity show an almost exactly periodic motion whose period equals that of the translational motion. Within each cycle of this motion are several oscillations, whose period is approximately that of the fundamental mode of torsional waves in the (entire) string. The largest peaks occur close to the onset and end of the rapid phase --- the passage of the Helmholtz corner. For the results in Figure 3, the bowing point is close to the measurement point, so the end points of the slow phase correspond, to within 2\% of a period, to those of the stick phase. The peak value of $r\omega$ is about one third of the speed of the string in the stick phase. So, in answer to one of the questions posed in this study, the torsional component of the velocity of the bow-string contact point is far from negligible.\\

The frequency domain representations of the torsional waves show a series of harmonics, whose (missing) fundamental is that of the \emph{translational} wave. There are also formants at frequencies corresponding to the fundamental and higher harmonics of the torsional standing waves on the (entire) string. In other words, the harmonics of the translational fundamental that are closest to the torsional resonances are much stronger than the others. Figure \ref{fig3}d shows that the 40 Hz component --- the translational fundamental --- is zero in the torsional signal. The presence of the many harmonics of this frequency give the signal a clear period of $\frac{1}{40}$ s, which is evident in the time domain representation (Figure \ref{fig3}c). This missing fundamental is clearly heard in the sound files, which are discussed below.\\

Together, these results show that the torsional wave has a period equal to that of the translational wave or equivalently that the frequency components of the torsional wave are driven at harmonics of the translational fundamental. This locking is explained by the much lower Q (more than fifty times lower) of the torsional waves. The time resolution of the measurements was 0.1 ms, or 0.004 of a typical value of a period. At this resolution, the jitter was not measurable. This small upper bound to the jitter is consistent with the common observation that the sound of a bowed string is musical, pleasant and has a well defined pitch.\\

The oscillations observed in the slow phase of the translational motion have a frequency similar to that of the torsional fundamental. It is therefore interesting to ask whether, during this phase, the centre of the string is moving at a constant speed plus $r\omega$. Unfortunately, the low pass filtering of the translational wave signal mentioned above makes it impossible to answer this question quantitatively. (A rectangular wave with high harmonics removed looks like a square wave with high harmonics added, in opposite phase.) However, the sum of the measured $v + r\omega$ shows oscillations with smaller amplitude than those in the $v$ signal alone (data not shown).\\

Figures \ref{fig4} and \ref{fig5} show the waves produced by bowing much closer to the bridge, at $\frac{L}{13}$. Bowing closer to one end of the string increases the relative power distributed to higher harmonics. This well known phenomenon is due in part to the requirement by the player to increase the ratio of normal force to bow speed so as to maintain bowing conditions within the 'Shelleng triangle' [4,5]. On the transverse wave, this effect is reduced by the low pass filtering discussed above. On the torsional wave, however, the effect is rather large, and so the second formant is several times larger than the first. The magnitude of the torsional waves is a little smaller in this case than in Fig 3, while the magnitude of the transverse waves is a little more, so the peak value of $r\omega$ is a smaller fraction of the transverse velocity.\\

Figures \ref{fig4} and \ref{fig5} also show the effect of tuning the string. The frequencies of the transverse waves in the examples shown are 40 and 50 Hz respectively. In separate experiments, the translational and torsional fundamental frequencies were measured. The tension that gave a translational fundamental of 40 Hz produced the first two torsional resonances at $240\pm 2$ Hz and $472\pm 3$ Hz.  The tension that gave a translational fundamental of 50 Hz produced the first two torsional resonances at $222\pm 2$ Hz and $438\pm 3$ Hz. The variation with tension of the frequencies of the torsional modes may be due to the effects of varying tension on the forces between the string's windings. There is a little hysteresis in this effect, which here may be related to the way in which individual windings slip over the bridge. This hysteresis leads to the uncertainties in the resonance frequency. The cause of the inharmonicity in the torsional resonances is unknown.\\

Different harmonics of the translational fundamental are therefore enhanced by the torsional resonances: when the translational fundamental is 40 Hz, its sixth and eleventh harmonics (240 and 440 Hz) are strongly present in the torsional signal. When the translational fundamental is 50 Hz, its fifth and ninth harmonics (250 and 450 Hz) are strongest.\\

It is interesting to listen to the measured signals $ v(t) $ and $ \omega(t) $, which we have stored as sound files at our web site \cite{15}. The spectrum of $ v(t) $ is expected to be rather different to that of the sound radiated by an instrument, because of the strong frequency dependence of the mechanical impedance at the bridge and the radiativity of the instrument \cite{16}. Nevertheless, the sound file of $ v(t) $ resembles somewhat a bowed bass playing E1. The sound of $ \omega(t) $ is also clearly identifiable as the sound of a low-pitched bowed string instrument, presumably because of the richness in harmonics and the transient during which periodic motion is established, but it does not sound like a bass. It has the pitch E1, the missing fundamental of the torsional wave, discussed above. Although E1 is the most noticeable pitch heard, the strong harmonic that falls near the torsional fundamental can be clearly heard as a distinct note whose pitch, of course, is an exact harmonic of the 40 Hz translational fundamental, as described above.\\

In all cases where the string was bowed by an expert, the translational and torsional waves are periodic waves with the same period, which is that of the fundamental of the translational wave for the string at that tension. In cases where the string was bowed by volunteers with no experience in bowing a string, the waves were often non-periodic, and the resultant sound unmusical (data not shown).\\

It is sometimes difficult to establish periodic motion quickly \cite{17} and indeed, steady periodic motion of the bowed string is only possible for a limited volume of the parameter space defined by the bow position, normal force applied, and bow speed [4]. It seems likely that the limits of this volume are determined by the requirement that self-sustained oscillations are produced with minimal jitter. The string player must maintain the operating point within that volume. Part of the skill in playing a string instrument presumably involves learning to establish quickly a periodic motion in which all of the waves, including the torsional waves, are forced into a harmonic relation.\\

\end{document}